\documentstyle[11pt,newpasp,twoside,epsf]{article}
\def\edcomment#1{\iffalse\marginpar{\raggedright\sl#1\/}\else\relax\fi}
\reversemarginpar

\begin{document}
\title{Intergalactic HI Clouds}
 \author{F.H. Briggs}
\affil{RSAA, Mount Stromlo Observatory, ANU\\
Cotter Road\\
Weston Creek 2611 ACT\\
Australia}

\begin{abstract}
Neutral intergalactic clouds are so greatly out numbered by galaxies that their 
integral HI content is negligible
in comparison to that contained in optically luminous galaxies. In fact,
no HI cloud that is not associated with a galaxy or grouping of galaxies 
has yet been identified. This points to a causal relationship that relies on
gravitational potentials that bind galaxies also being responsible for confining
HI clouds to sufficient
density that they can become self-shielding to the ionizing background
radiation. Unconfined clouds of low density become ionized, but confined
clouds find themselves vulnerable to instability and collapse, leading to
star formation.
\end{abstract}

\section{Introduction}

Astronomers think of hydrogen as the most primitive of the atomic elements.
Hydrogen, helium, and a trace of lithium, along with their isotopes, constitute
the primordial mix of baryonic material
emanating from The Big Bang origin of the Universe;
the heavier elements form over time through nuclear processes in the interiors of
successive generations of
stars.

A natural consequence has been the quest to identify remnant clouds of primordial
hydrogen in a pristine state. The suspicion a few decades ago was that
clouds of neutral atomic hydrogen might float freely in the intergalactic
medium, where they would exist as self-gravitationally bound systems in much
the same way that galaxies of stars are bound together. Subsequently, astronomers
have succeeded in identifying the ionized intergalactic medium through the
discovery of the Lyman-$\alpha$ forest, but they have yet to find free-floating
clouds of neutral gas.

This contribution to 
these proceedings gives a short review of the limits on intergalactic neutral
clouds, along with a discussion of why we no longer expect them to exist.
It is important to point out from the start that there are numerous 
HI clouds swarming around galaxies and groups. In one sense, one
could argue that these clouds ``between the galaxies'' might qualify
as intergalactic.
However, in all well-studied cases, these clouds appear  to be products of their
environments (possible tidal or interaction remnants), and their existence in
the neutral phase seems to be a consequence of confinement by the gravitational
potentials of the host galaxies or groups. HI clouds are not found unaccompanied
by galaxies.

\begin{figure}[t]
\hglue -0.4cm
\epsfxsize=14.57cm
\epsfbox{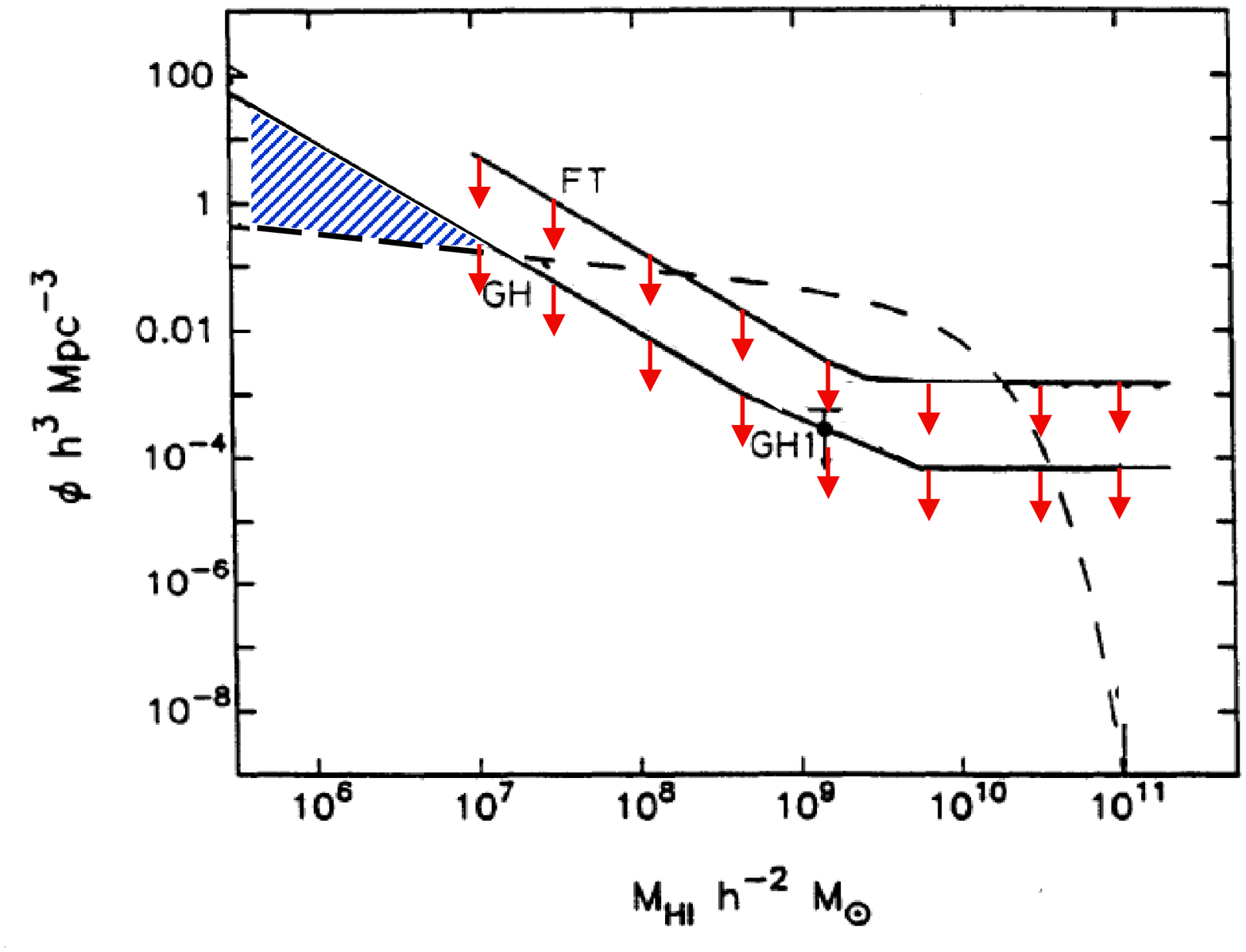}
\caption{The HI mass function $\phi(M_{HI})$ compared with
upper limits to the number densities of intergalactic HI clouds
(after Briggs 1990) per Mpc$^3$ per decade of HI mass.  
The upper limits derived by Fisher \& Tully (1981)
(upper solid curve with arrows) and Briggs (1990) (lower solid curve with
arrows) are compared to the HI mass function for galaxies (dashed curve)
as derived by
Briggs (1990) using the optical luminosity function of Felton (1985) and
the HI to optical luminosity ratio trend from Fisher \& Tully (1975). The
hatched region at left is the only zone where the number of intergalactic
clouds might have exceeded the number of galaxies of the same HI mass.
\label{himf.fig}
}
\end{figure}

\section{Limits to the numbers of intergalactic HI clouds}

A number of low redshift 21cm line surveys for intergalactic HI clouds 
took place in the 1970's, and the resulting trend of associating all HI signals
with optically luminous galaxies has continued to this day. Fisher and
Tully (1981) summarized the surveys at that time, and they concluded that
the mass in intergalactic clouds in the mass range $10^7$ to $10^{10}M_{\odot}$
is less than 6\% of the luminous mass in galaxies and less than 
$5{\times}10^{-4}$ the density required for a closed Universe. With 98\% of
their survey of 1787 nearby galaxies completed, they had found no convincing
detection of an HI signal without an optical counterpart. 

There is a clear
trend of increasing HI-mass to optical luminosity ratio $M_{HI}/L$ with
decreasing optical luminosity for late-type spiral and irregular galaxies
(Fisher \& Tully 1975). The early survey by Lo \& Sargent (1979) discovered 
several dwarf galaxies in nearby groups, and some of these had $M_{HI}/L>10$.
One might argue that these could be called ``HI clouds with a trace of
stellar contamination,'' although in these cases, the optical observations
indicate dwarf galaxies of ordinary appearance... they are simply extremely
rich in HI (Warren, Jerjen et al, this symposium).

Briggs (1990) revisited the upper limits to intergalactic clouds and compared
them to the number density of HI-rich galaxies by deriving $\phi(M_{HI})$
the HI mass function
(HIMF) from the properties of the optically identified galaxy populations.
Felton's (1985) optical luminosity function (with faint-end power law
exponent of $\alpha=-1.25$ and the $M_{HI}/L(L)$ trend from Fisher and Tully
(1975) led to the HIMF shown in Fig.~1. In the mass range around
$M_{HI}{\sim}10^9M_{\odot}$, the Fisher-Tully limits constrain
the intergalactic HI clouds to be less numerous than galaxies by a factor
of ten.  In addition, 
Briggs reviewed the 21cm galaxy surveys made in the decade since
Fisher and Tullys' catalog. The important Arecibo surveys of Giovanelli and Haynes
and colleagues (1989a) had turned up the ``Virgo Cloud''(Giovanelli \& Haynes 1989b), 
a Magellanic irregular
galaxy that had been mistaken for an intergalactic cloud, and the confidence
of the detection of this object
with $M_{HI}{\sim}10^9M_{\odot}$ in an Arecibo calibration scan
 reinforced the idea that the Arecibo surveys
could place strict constraints on the cloud population. These types of surveys, which
are designed to study optically selected galaxies, rely on a calibration
scan from blank sky to go with every observation with the telescope aimed 
at a galaxy. The number of objects discovered by chance in the off-source
scans provides a measure of the intergalactic cloud detection rate and places 
upper limits on their space density. The upper limits then become quite
strict over the mass range $10^8$ to $10^{10.2}M_{\odot}$, constraining the
population to be a factor of 10 to 100 less numerous than galaxies of similar
mass over this range.

Only in the mass regime $M_{HI}<10^7M_{\odot}$ do the upper limits permit
HI clouds to outnumber galaxies. The upper limits to space density still
constrain the integral mass density of such low mass clouds to be negligible
in comparison to the mass content of the massive, luminous galaxies that
are the dominant reservoirs of atomic hydrogen at the present age of the
Universe.

The integral HI content of the Universe at present is computed by the integral
$\rho_{HI}=\int M_{MI}\;\phi(M_{HI})\;dM_{HI}$.  The HIMF from
Briggs (1990) yields a value for $\rho_{HI}$
of $7{\times}10^7M_{HI}$Mpc$^{-3}$ with a faint-end HIMF slope of
$\alpha=-1.28$ ($H_o=75$~km~s$^{-1}$Mpc$^{-1}$).  
Recent values from the Parkes HIPASS survey (Zwaan et al 2003) are
$\rho_{HI}=6.1{\times}10^7M_{HI}$Mpc$^{-3}$ with $\alpha=-1.3$, indicating a
convergence between measurements based on optical and radio selection
techniques.

A comprehensive understanding of 
the Galactic High Velocity Cloud population remains illusive.
Recently, Blitz et al (1999) and  Braun and Burton (1999)
have explored the possibility that
the in-falling HVC population of HI clouds  are remnants of a primordial
extragalactic population of dark matter mini-halos. 
In this scenario, the HI masses of the clouds would typically
be larger than ${\sim}$10$^7M_{\odot}$, and every large galaxy should be surrounded
by a similar halo of a few hundred of these objects, assuming the phenomenon is a genuine
and common feature of galaxy formation and evolution.  The fact that nearby galaxies
and groups do not possess such a halo of small clouds (Zwaan \& Briggs 2000, Zwaan
2001) has ruled out this idea, requiring that the clouds must be at least
an order of magnitude
less massive and fall at distances within ${\sim}$200~kpc of the Milky Way, well
within our Galaxy's halo.

The clear association of neutral gas clouds with star-bearing galaxies implies that
the HI relies on the confinement of the galaxies' gravitational potentials for
their survival.

\section{What happened to the HI clouds that formed at Recombination?}

Figure~2 gives an overview of the history of neutral gas clouds
over the age of the Universe.  It begins at the phase transition
corresponding to the release of the Cosmic Microwave Background
(at $z\sim 1100$), when the ionized baryons and elections
combine to become a neutral gas that is transparent to the CMB photons.
There follows the only period, lasting
about 100 million years, when the majority of the Universe's atoms
are neutral. This period, known as the `Dark Age,' ends when the first
objects collapse as a result of gravitational instability, providing sources of
ionizing energy. We refer to the end of the Dark Age as the `Epoch of
Reionization' (EoR), and
we associate the EoR with the onset of the first
generation of stars (which form in the most over-dense regions) and the
appearance of protogalactic objects, which become the building blocks of
galaxies.
\begin{figure}[t]
\epsfxsize=11.7cm
\hglue 0.8cm\epsfbox{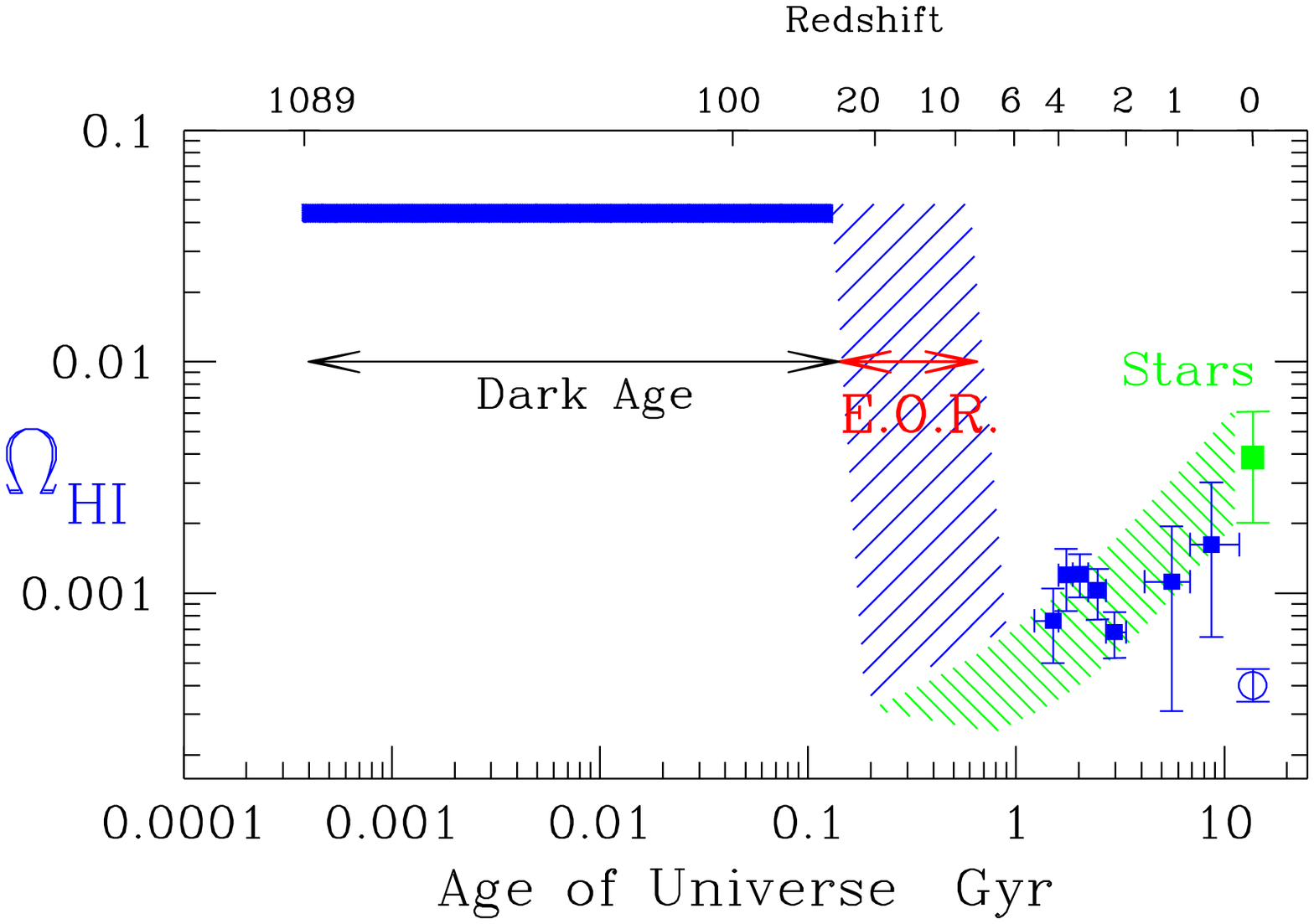}
\caption{History of the neutral hydrogen content of the Universe.
The logarithm of the neutral gas density normalized to the `closure density'
necessary to close the Universe is plotted as a function of the age of the
Universe. Square filled points are measurements from Damped Lyman-$\alpha$
QSO absorption-line statistics. The open circle at far right represents the
neutral gas content of the present day ($z=0$) Universe. For comparison,
the rising trend of stellar mass content appears has a hatched envelope, which
increases to the value measured at $z=0$ from the optical luminosity density
of stars.
 \label{omeg.gif}}
\end{figure}

In the diagram of Fig.~2, the EoR is also marked by
the appearance of a second shaded region that indicates schematically
the beginnings of the build up
of mass in stars, as subsequent generations of star formation gradually lock
increasing numbers of baryons into low mass, long lived stars. The stellar mass
content of the Universe rises steadily from the EoR to the present, where we have
precise measurements through meticulous inventories of the numbers of galaxies and
their luminosities (i.e., the galaxy luminosity function and the integral
luminosity density)(see for instance, Madgwick et al 2002). The neutral hydrogen
content at $z\approx 0$ is measured by 21cm line surveys 
(Zwaan et al 1997, Zwaan et al 2003)
such as those described in the previous section.

Through the period following the EoR, astronomers have
statistical measures of the HI content as a
function of time through the observation of QSO absorption lines. 
Any gas rich object
that populates the Universe has a random chance of intervening along the line 
sight to distant objects.   Quasi-stellar objects are especially useful
as background sources since
they have strong optical and UV continuum emission against which intervening
gas clouds can imprint a distinctive absorption line spectrum.  In the case of
thick clouds of neutral gas, the Lyman-$\alpha$ line of HI is so strong that
it presents an easily recognized `damping wing' profile, which has led to
the Damped Lyman-$\alpha$ (DLA) class of QSO absorption line (Wolfe et al 1986);
in the minds of most astronomers, the DLAs are
associated with gas-rich protogalaxies, which
are the precursors of the larger galaxies that we observe around us at present
(Prochaska \& Wolfe 1997, Haenelt et al 1998).

\subsection{The ionization/recombination competition}

Once sources of ionization formed in the earliest
astrophysical structures, the survival of neutral clouds became
a competition between ionization and recombination rates.
Since ionization is such a common hazard to the existence of
neutral atoms, it is natural to ask, ``how rapidly can an ion
recover through recombination, after it has chanced to become ionized?''
For hydrogen, the recombination rate $R$ is easily computed (for instance
Spitzer 1978), and the time $t_{recomb}$ it takes for recombination to eliminate
the electrons in a cloud of electron density $n_e$ is
\begin{equation}
t_{recomb}=\frac{n_e}{R} = \frac{n_e}{n_en_p\alpha_{recomb}}
=\frac{1}{n_p\alpha_{recomb}}\approx\frac{T^{1/2}}{2{\times}10^{-11}n_p} {\rm \ \ sec}
\end{equation}
where $n_p$ is the proton density and $\alpha_{recomb}$ is the recombination
coefficient. To get a feeling for the vulnerability of the bulk of the
baryons that populate the intergalactic medium, the number density of
baryons $n_{baryon}$ forms an estimate of $n_p$; over-dense regions will
have relatively shorter recombination times. In an expanding Universe,
$n_p\sim n_{baryon}\sim (1+z)^3$, so that
\begin{equation}
t_{recomb} \propto \frac{T^{1/2}}{(1+z)^3}
\label{recomb_time.eqn}
\end{equation}
\begin{figure}[t]
\vglue -.5cm
\hglue 1.0cm\epsfxsize=13.cm
\epsfbox[150 420 500 690]{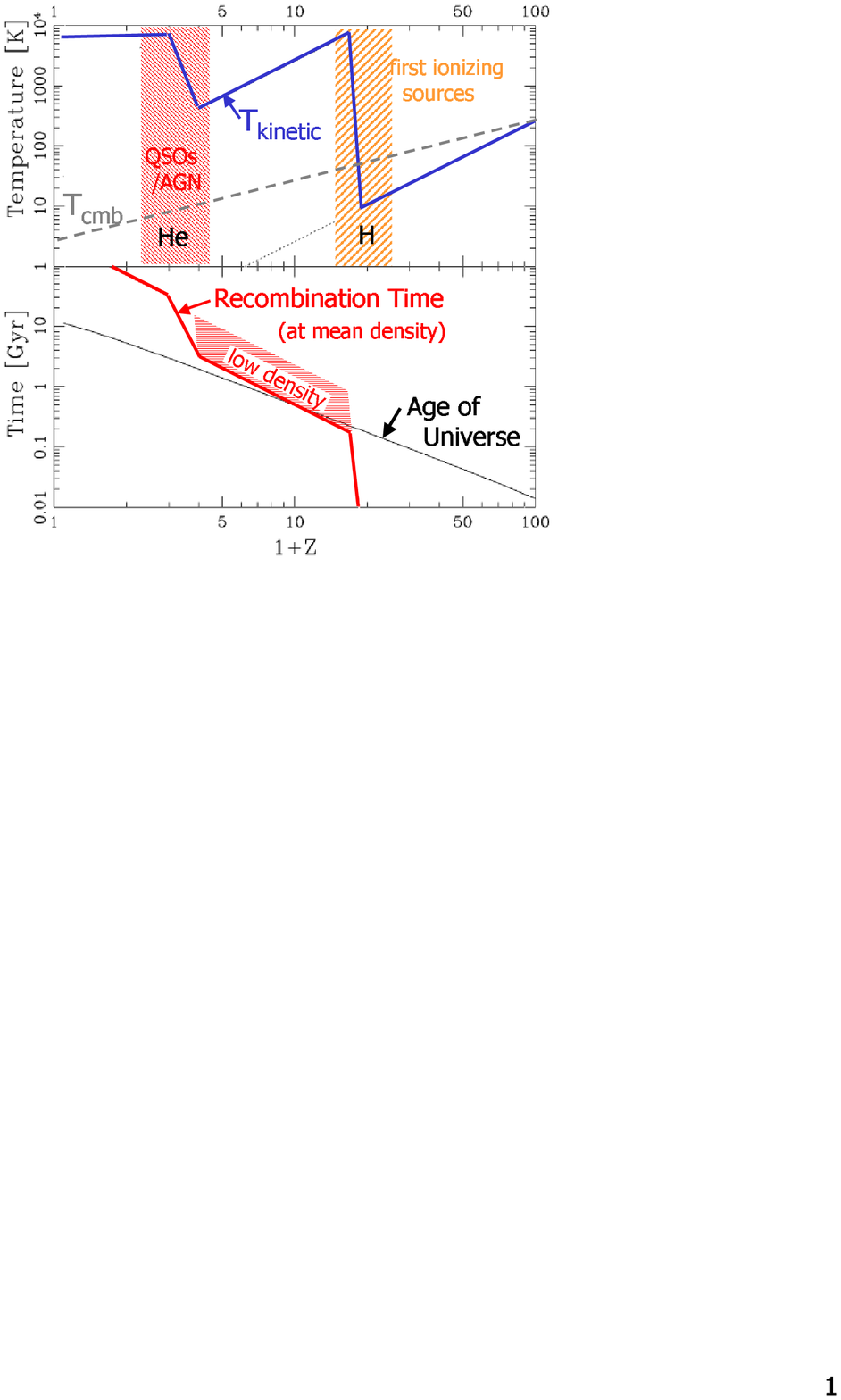}
\caption{Recombination time in the intergalactic medium as a function of
redshift $z$. {\it Upper Panel:} Kinetic temperature $T_k$ and CMB
temperature $T_{CMB}$ vs. redshift. Episodes of heating through
photoionization of hydrogen occur during the Epoch of Reionization
and during the reionization of helium at a later time by the harder
radiation from active galactic nuclei. {\it Lower Panel:} Recombination
Time for an intergalactic medium of mean baryonic density,
compared with the Age of the Universe as a function of redshift.
\label{recomb_time.fig}}
\end{figure}
The recombination time of the IGM at mean density has a strong dependence
on age of the Universe through the $(1+z)^3$, and a modest dependence
on temperature $T$. Fig.~3 provides a rough
illustration of how the IGM temperature varies with time and the
net influence of
the dependencies in Eqn.~2
on the ionization state of the Universe.

If the expansion of the Universe would allow a completely uniform
expansion of the IGM without the growth of gravitationally-driven
density instabilities, the gas kinetic temperature would decline
in the adiabatic expansion with dependence $T_k\propto (1+z)^2$.
At the same time, the CMB radiation temperature declines as
$T_{CMB}\propto (1+z)$,
causing the two temperatures to decouple after $z\approx 100$, when
electron scattering ceases to be effective. The IGM is reheated when 
photoionization spreads through the medium generating energetic photoelectrons
that deposit their kinetic energy through scattering. Once the IGM is
fully ionized, there is no effective means of adding energy to the
gas, since the photons generated by the stars can now flow uninhibited
through a transparent medium, and the IGM again cools adiabatically due to
Universal expansion.

A similar heating event can occur during the age around $z\sim 2$
 when QSOs are most common.  QSOs, as well as lesser AGN, radiate
photons that are capable of ionizing helium, and these
harder photons generate photoelectrons throughout the IGM, providing
a second round of localized heating.

The two heating events impact on the ability of the Universe to
recombine. The lower panel of Fig.~3 compares the
recombination time $t_{recomb}$ of an IGM of mean density to the
age of the Universe $t_{age}$
as a function of redshift. If $t_{recomb}$ is long compared to $t_{age}$,
the IGM would never recover from its ionized state, even if the source
of ionizing photons were turned off completely. The figure shows that there
is period between the two heating events, when recombination
can compete with ionization, depending on 1) the intensity of the ionizing
flux and 2) the local density. Under-dense regions would already be
destined to stay forever ionized. Over-densities, especially those clouds
confined in gravitational potential wells, may be able to recombine.

At low redshifts, the density of the mean IGM has become so dilute, that
the IGM will remain ionized, even though the photoionizing background
from AGN tails off.

The existence of atomic hydrogen clouds at all at low redshift in and around
galaxies is due
to their confinement to high density (greater than ${\sim}0.1$~cm$^{-3}$)
where the recombination times are $<10^5$yrs, and recombination can
compete effectively to make self-shielding clouds.

\section{Summary}

There is no evidence that truly intergalactic neutral hydrogen clouds exist
at the present age of the Universe. The observational evidence constrains them to
constitute a small fraction of the neutral gas contained in galaxies and
to be at most a tiny fraction of the Universal baryons. The bulk of the baryons
appear to be contained in the ionized intergalactic medium and stars. Where the
hydrogen is neutral, it is confined in galactic potential wells, where  1) it is
self-shielding against ionizing radiation by having sufficient density that
recombination can compete with the rate ionization and 2) it is
vulnerable to the instabilities that lead to star formation. 

There is a limited range of options for neutral gas in a dark matter halo.
The confining potential of 
a low mass halo (with virial temperature below ${\sim}10^4$K) that has formed early
as part of the first star forming protogalaxies will be unable to retain its 
gas during the first bursts of star formation, and the halo will be evacuated.
Low mass halos that form later (after reionization and reheating), will be unable
to capture and confine hot baryons from the intergalactic medium, and they therefore
are destined to remain free of neutral gas.  Massive halos can retain and capture
gas, but, once confined in a halo, the gas is susceptible to collapse and star formation.
The strong association of HI with star bearing halos makes sense.

In fact, there is a further consequence of this apparently complete 
correlation:  Since all neutral gas
appears to be associated in some way with stellar light, this constitutes evidence
that there are no dark Dark Matter halos that are comparably massive to the Dark Matter
halos that host galaxies, since such halos would be able to compress gas clouds
in much the same way that the outer disks and rings of material hover in the
outer regions of late-type galaxies (Schneider et al 1989, van Driel \& van Woerden
1991, Oosterloo et al 2003). Since rings or isolated clouds without stars are not
observed, only a small minority of halos of galaxy-sized mass that are devoid of stellar
emission can exist.

\acknowledgments{The author is grateful to the organizers of the Symposium
for assembling a fascinating meeting and to the editors of the proceedings
for their patience.}

\end{document}